\documentstyle[12pt,epsf]{article}

\advance\voffset by -1.5cm
\advance\hoffset by -1.5cm
\def\be{\begin{equation}}
\def\ee{\end{equation}}

\def\pmb#1{\setbox0=\hbox{#1}%
 \kern-.025em\copy0\kern-\wd0
 \kern.05em\copy0\kern-\wd0
 \kern-.025em\raise.0433em\box0 }
\def\A{{\cal A}}

\def\3{\ss}
\def\sq{\hbox{\rlap{$\sqcap$}$\sqcup$}}
\def\qed{\ifmmode\sq\else{\unskip\nobreak\hfil
\penalty50\hskip1em\null\nobreak\hfil\sq
\parfillskip=0pt\finalhyphendemerits=0\endgraf}\fi}
\def\half {\frac{1}{2}}

\def\bbbz {{\sf Z\!\!Z}}

\def\Nu{{{\cal N}}}
\newcommand{\ket}[1]{|#1\rangle}
\newcommand{\bra}[1]{\langle#1|}
\def\Tr{{\rm Tr}}

\begin{document}

\thispagestyle{empty}
\def\thefootnote{\fnsymbol{footnote}}
\begin{flushright}
  hep-th/9705130\\
  HUTP-97/A022 \\
  BRX TH-414\\
  PUPT-1703\\
\end{flushright}
\vskip 0.5cm

\begin{center}\LARGE
{\bf Branes, Orientifolds and the Creation of Elementary Strings}
\end{center}
\vskip 0.5cm
\begin{center}\large
       Oren Bergman%
       $^{a,b}$\footnote{E-mail  address: {\tt
bergman@string.harvard.edu}},
       Matthias R. Gaberdiel
       $^{b}$\footnote{E-mail  address: {\tt
gaberd@string.harvard.edu}} and
       Gilad Lifschytz%
       $^{c}$\footnote{E-mail  address: {\tt
       gilad@puhep1.princeton.edu}}
\end{center}
\vskip0.5cm

\begin{center}\it$^a$
Department of Physics \\
Brandeis University \\
Waltham, MA 02254
\end{center}

\begin{center}\it$^b$
Lyman Laboratory of Physics\\
Harvard University\\
Cambridge, MA 02138
\end{center}

\begin{center}\it$^c$
Joseph Henry Laboratories\\
Princeton University\\
Princeton, NJ 08544
\end{center}

\begin{center}
May 1997
\end{center}

\begin{abstract}
The potential of a configuration of two Dirichlet branes for which the
number of ND-directions is eight is determined. Depending on whether
one of the branes is an anti-brane or a brane, the potential vanishes
or is twice as large as the dilaton-gravitational potential. This is
shown to be related to the fact that a fundamental string is created
when two such branes cross. Special emphasis is given to the D0-D8
system, for which an interpretation of these results in terms of the
massive IIA supergravity is presented. It is also shown that the
branes cannot move non-adiabatically in the transverse direction.
The configuration of a zero brane and an orientifold $8$-plane is
analyzed in a similar way, and the implications for the type IA-heterotic
duality and the heterotic matrix theory are discussed.
\end{abstract}

\vfill
\setcounter{footnote}{0}
\def\thefootnote{\arabic{footnote}}
\newpage

\renewcommand{\theequation}{\thesection.\arabic{equation}}

\section{Introduction}
\setcounter{equation}{0}

It has become apparent in the last two years that Dirichlet-branes
(D-branes) \cite{dlp,pol} are the ``solitonic'' objects of string
theory which are relevant for the non-perturbative regime of the
theory \cite{ht,wit}. These D-branes can be studied in a variety of
ways, and this has proved very useful. For example, D-branes can be
analyzed using techniques of perturbative string theory, either in
terms of suitable boundary conditions of open strings, or as coherent
states in the closed string sector. It is also possible to study the
field theory on the world-volume of the D-brane, and D-branes can be
analyzed from the point of view of the low-energy supergravity
theory. Many calculations have been performed using different
techniques, and this has allowed for a number of consistency checks.

In this paper we shall analyze a certain class of configurations of
two D-branes which is characterized by the property that the number of
mixed boundary conditions of an open string stretching between
them is ND$=8$. It is well known that a system of two D-branes retains
unbroken supersymmetries if ND$=0,4$ or $8$ \cite{psj}. The cases
ND$=0$ and $4$ have been analyzed from a number of different points of
view \cite{pol,gil,dkps}, and it has been found that two such branes
do not exert a force onto each other. Furthermore, there exist
solutions to the supergravity equations which reproduce all of these
configurations. On the other hand, the case ND$=8$ is much less
understood, and there exist contradictory claims in the literature.
For example
\begin{itemize}
\item The string calculation predicts that the force vanishes for
ND$=8$ configurations \cite{gregut,gil,bergmangaberd}. In the open
string description there is a contribution from the sector R$(-1)^F$,
which is absent for all other D-brane configurations. From the closed
string point of view, the cancellation occurs between the NS-NS and the
R-R sector. The contribution in the R-R sector is somewhat puzzling
as one does not expect, for example, a D0-brane and an D8-brane to
interact through a R-R gauge field.

\item There are solutions of the supergravity equations of motion which
preserve a quarter of the supersymmetries and look like the brane
configurations of ND$=8$ \cite{gkt}. For the case of the D0-D8
system, however, there exists no solution.

\item A Yang Mills analysis of the D0-D8 system in
\cite{df} shows that there is a non-zero force between D0-branes
and D8-branes, and a non-zero force between D0-branes and
an orientifold-8-plane ($\Omega8$-plane). The forces cancel if eight
D8-branes (and their images) are at the orientifold plane. The system
in \cite{df} was used successfully to explain certain aspects of the
$E_{8}\times E_{8}$ heterotic matrix model \cite{ks,kimray,lowe}.
These results correspond to the string calculation if only the NS-NS
sector contribution is taken into account.

\item From the point of view of the supergravity, the force on the
D$p$-brane in the background of a D$p'$-brane, where $p'>p$, can be
calculated. For ND$=8$, the force which is due to the exchange of the
graviton and the dilaton is repulsive.

\item A careful treatment of the D0-D8 system using massive
supergravity theory shows that this system is inconsistent
unless extra macroscopic elementary strings are introduced that end
on the D0-brane \cite{polstrom}.
\end{itemize}

We will show that indeed in the case of ND$=8$ the string calculation is
correct. The Yang-Mills analysis correctly accounts for the short
distance degrees of freedom, but misses a linear potential coming from
the R-R sector. For the case of the D0-D8 system, this potential can
be interpreted as coming from an elementary string, which is thus in
agreement with \cite{polstrom}. The calculation also indicates that an
elementary string is created when the two branes cross each other
adiabatically, and this is related to the Hanany-Witten effect
\cite{hw} by a series of dualities.
\smallskip

We analyze the velocity dependent force in the D0-D8 system, and
we find, somewhat surprisingly, that the branes cannot move in the
transverse direction non-adiabatically.  This has important implications
for the non-perturbative behavior of type IIA string theory in the
presence
of D8-branes: the theory appears to be ten-dimensional at strong
coupling.  This is consistent with the fact that there does not exist
a massive eleven-dimensional supergravity \cite{Deser}.

The paper is organized as follows. In section~2, we explain the open
string theory calculation, and show that a fundamental string is
created when the two branes cross; we also relate this to the
Hanany-Witten effect. In section~3, we explain how the supergravity
analysis of Polchinski and Strominger \cite{polstrom} gives an
interpretation for the R-R contribution force in the D0-D8 case. In
section~4, we analyze the configuration of the D0-brane and the
$\Omega 8$-plane, and in section~5, we determine the
velocity dependent potential for the D0-D8 and the D0-$\Omega$8
configurations. In section~6 we show that our results are consistent
with the world-line theory point of view, and we explain some of the
implications for the  matrix models.
We have included an appendix, where the results are
derived from a closed string point of view.
\smallskip

While this paper was being finalized, the papers \cite{bdg,dfk,jpierre}
appeared in which some overlapping results have been obtained.

\section{Stationary Potential in the ND$=8$ system}
\setcounter{equation}{0}

Let us consider a stationary configuration of two parallel or
orthogonal D-branes, and suppose that we have chosen the coordinates
of our spacetime so that the branes are parallel to the coordinate
axes. Consider an open string that is stretched between the two
branes. This string satisfies for every coordinate a Neumann (N) or
Dirichlet (D) boundary condition at either end. We denote by NN
the number of coordinates for which both ends have a N condition, by
DD the number of coordinates for which both ends satisfy a D
condition, and by ND the number of coordinates for which one end has a
D, and one end a N condition. As we are considering the
ten-dimensional superstring, NN$+$ND$+$DD$=10$. We shall always assume
that the time direction $x^0$ is a NN direction, so that we are
considering D-branes rather than D-instantons. The D-branes will be
separated along the DD directions by a vector ${\bf R}$, and the free
energy of such a configuration is given at one loop by the annulus
amplitude
\be
\A = 2 V_{NN} \int {d^{NN}k \over (2\pi)^{NN}}  \int_0^\infty
{dt\over 2t} \Tr \bigg[
        e^{-2\pi\alpha^\prime t(k^2 + M^2)}
        (-1)^{F_S}  {1\over 2} (1+(-1)^F)
       \bigg] \,,
\label{openamplitude}
\ee
where $V_{NN}$ is the spacetime volume of the NN-directions,
$F_S$ and $F$ are the spacetime and worldsheet fermion
numbers, respectively, and
\be
M^2 = {R^2\over 4\pi^2\alpha^{\prime 2}} + {1\over\alpha^\prime}
            \Bigl\{\sum_n (\alpha_{-n} \cdot \alpha_n + nb_{-n}c_n +
                     nc_{-n}b_n)
            + \sum_m m(\psi_{-m}\cdot \psi_m + \beta_{-m}\gamma_m -
                     \gamma_{-m}\beta_m ) + a  \Bigr\} \,.
\ee
Here $a=0$ in the Ramond (R) sector, and $a=-1/2+\hbox{ND}/8$ in the
Neveu-Schwarz (NS) sector. The moding of the bosonic and fermionic
oscillators ($\alpha_n^\mu, \psi_m^\mu$) in the NN and DD directions
is given by
\be
 n\in \bbbz \quad , \quad
 m \in \left\{ \begin{array}{ll}
                       \bbbz & \mbox{R} \\
                       \bbbz + 1/2 & \mbox{NS}
                      \end{array}
          \right. \,,
\label{NNmoding}
\end{equation}
and in the ND directions by
\begin{equation}
 n\in \bbbz + 1/2 \quad , \quad
 m \in \left\{ \begin{array}{ll}
                       \bbbz +1/2 & \mbox{R} \\
                       \bbbz  & \mbox{NS}
                      \end{array}
          \right. \,.
\label{NDmoding}
\ee
The moding of the ghost and superghost oscillators
is always as in (\ref{NNmoding}), so that their contribution will
cancel against the bosonic and fermionic contributions in two NN or DD
directions.

Let us concentrate on the case D$p\perp$D$(8-p)$
(ND$=8$). This includes the D-particle
D8-brane system, the configuration of a D7-brane and an orthogonal
D-string, {\it etc}. Let us denote by $x^9$ the transverse DD
direction, and let $R$ be the transverse distance between the two
branes along $x^9$. The potential between the branes is then
\be
V_{Dp\perp D(8-p)} (R)  = -2 \int {dk_0 \over 2\pi}  \int_0^\infty
{dt\over 2t} \Tr \bigg[
        e^{-2\pi\alpha^\prime t(k_0^2 + M^2)}
        (-1)^{F_S}  {1\over 2} (1+(-1)^F)
       \bigg] \,.
\label{08openamp}
\ee
Integrating over $k_0$ and performing the traces gives \cite{gil}
\begin{eqnarray}
V_{Dp\perp D(8-p)} (R) &=& -\int {dt\over 2t} (8\pi^2\alpha^\prime
t)^{-1/2}
         e^{-R^2 t/(2\pi\alpha^\prime)}  \bigg[
         {f_2^8(q) - f_3^8(q) \pm f_4^8(q)
          \over f_4^8(q)} \bigg] \nonumber \\
    &=& -{1\over 2} T_0 R \Big[ 1 \mp 1 \Big]  \,,
\label{08amp}
\end{eqnarray}
where $q=e^{-\pi t}$, and the functions $f_i$ are defined in appendix~A.
Here $T_0 \equiv (2\pi\alpha^\prime)^{-1}$ is
the string tension, and we have used the ``abstruse identity''.
The three terms in the bracket correspond to the traces over the
NS, R and R$(-1)^F$ sector, respectively, and the sign of the third
term corresponds to the choice of the action of $(-1)^F$ on the
R-sector ground states. The trace over NS$(-1)^F$ vanishes
identically, as the fermionic oscillators in the transverse directions
$\mu=1, \ldots, 8$ have zero modes. The R-sector, on the other hand,
has only fermionic zero modes in the $\mu=0,9$ directions; their
contribution cancels against the (bosonic) superghost zero modes, and
the R$(-1)^F$ trace does contribute.\footnote{This is actually true for
any system with NN$+$DD$=2$, in particular ND$=8$, but also for boundary
conditions involving background worldvolume gauge fields for
$\mu=1,\ldots,8$.} The contribution of this sector is
independent of $q$, implying that only the massless string modes, rather
than the full string spectrum, contribute.

The open string one-loop calculation can be related, by a modular
transformation, to a closed string tree-level calculation (see, for
example, appendix~A). Under this transformation, the NS and R
contributions come from the (closed) NS-NS sector, and the R$(-1)^F$
contribution from the (closed) R-R sector. The former represents
(in the large $R$ limit) the combined interaction of the graviton and
the dilaton which is repulsive in this case.\footnote{This combined
interaction is attractive for ND$<4$, and repulsive for ND$>4$; it
vanishes for ND$=4$.} The sign of the latter contribution depends on
whether we are considering two branes or two anti-branes, or one brane
and one anti-brane.

It follows from (\ref{08amp}) that the vacuum energy of a brane-brane
system differs from that of the corresponding brane-anti-brane
system. In one case the vacuum energy vanishes, whereas in the other
it is equal to $-T_0 R$. Let us now consider the effect of rotating
the D$(8-p)$-brane by $\pi$ in the plane spanned by a direction of the
D$(8-p)$-brane and the transverse direction $x^9$; this turns the
D$(8-p)$-brane into an anti-brane ($\overline{\mbox{D}(8-p)}$-brane).
On the other hand, this operation is topologically equivalent to
moving the D$p$-brane across to the other side of the the
D$(8-p)$-brane, and we conclude that the sign in (\ref{08amp}) flips
as the D$p$-brane moves from one side to the other.

Suppose that the system is in the vacuum state of vanishing energy
when the D$p$-brane is to the left of the D$(8-p)$-brane. As we move
the D$p$-brane adiabatically to the right across the D$(8-p)$-brane,
the system remains in the same state. As explained above however, the
vacuum on the right hand side is different. The difference in energy
is manifested in the creation of a string between the branes (see
figure~1). 

\begin{figure}[htb]
\epsfysize=3cm
\centerline{\epsffile{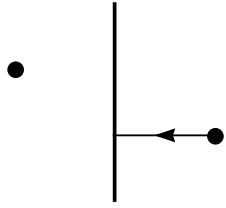}}
\caption{A D$p$-brane crossing a D$(8-p)$-brane.}
\end{figure}

The force that is felt by the D$p$-brane in the presence of a
D$(8-p)$-brane depends on the actual energy of the system, rather than
its vacuum energy. In the absence of strings, the force is zero
on one side and constant repulsive $-T_0$ on the other. On the other
hand, as we have seen above, if we start in a state without any
strings and with vanishing force, then a string is created as the
branes cross and the force will remain zero.
\smallskip

The process of string creation is equivalent to the effect of
Hanany and Witten \cite{hw}, as can be seen from the following chain
of dualities. We start with the configuration of \cite{hw}
\begin{eqnarray}
 NS5 &:& (x^{0},x^{1},x^{2},x^{3},x^{4},x^{5} )\nonumber \\
  D5 &:& (x^{0},x^{1},x^{2},x^{7},x^{8},x^{9} )\nonumber \\
  D3 &:& (x^0,x^{1},x^{2},x^6) \,.
\end{eqnarray}
Here each 5-brane induces on the other a charge which is
equal to $\half$ of the charge carried by the end of D3-brane.
We T-dualize along $(x^1, x^2)$, thereby mapping the system to
\begin{eqnarray}
 NS5 &:& (x^{0},x^{1},x^{2},x^{3},x^{4},x^{5}) \nonumber \\
  D3 &:& (x^{0},x^{7},x^{8},x^{9}) \nonumber \\
  D1 &:& (x^0,x^6) \,.
\end{eqnarray}
We then perform a IIB S-duality transformation, giving
\begin{eqnarray}
  D5 &:& (x^{0},x^{1},x^{2},x^{3},x^{4},x^{5}) \nonumber \\
  D3 &:& (x^{0},x^{7},x^{8},x^{9} )\nonumber \\
  F1 &:& (x^0,x^6 )\,.
\end{eqnarray}
This is a particular ND$=8$ configuration; all the others are obtained
by T-duality. In particular, the D0-D8 system is obtained by
T-dualizing along $(x^{7},x^{8},x^{9})$.

The analogy with the Hanany-Witten effect extends also to the induced
charges on the D-brane world-volume, which can be derived from the
WZW-terms in the world-volume theories. The conservation of these
charges requires that a string of appropriate orientation is either
created or annihilated when the branes cross. On the other hand,
it is not yet clear what accounts for the R$(-1)^F$ contribution to
the energy. In the following section we shall see that for the special
case of the D0-D8 system further constraints give a clearer picture.

\section{The D0-D8 System Re-examined}
\setcounter{equation}{0}

Among the ND$=8$ configurations the D0-D8 system is special in two
respects:
\begin{list}{(\roman{enumi})}{\usecounter{enumi}}
\item The D0-brane cannot support a world-volume charge.
\item The low-energy effective theory is {\it massive} IIA
supergravity \cite{roman,pol}, rather than massless IIA or IIB
supergravity.
\end{list}
The first point implies that a fundamental string cannot end on an
isolated D0-brane \cite{Strom}. On the other hand, as pointed out in
\cite{polstrom}, because of (ii) fundamental strings must end on
a D0-brane in the presence of D8-branes. Let us briefly review the
argument.
The action of massive IIA supergravity contains the term
\be
 \int d^{10} x \sqrt{-g} e^{3\phi/2} [ dA^{(1)} + mB^{(2)} ]^2 \,,
\label{higgs}
\ee
which is a generalization of the Higgs mechanism, in which the two-form
field $B^{(2)}$ acquires mass $m$, and $A^{(1)}$ plays the role of the
Goldstone boson. In the background of D8-branes $m\neq 0$, and the
equation of motion for $B^{(2)}$ gets a contribution from
(\ref{higgs})
\begin{equation}
d \ast (e^{-2\phi} dB^{(2)}/2)=m \ast(dA^{(1)}+mB^{(2)}) \,.
\label{msg}
\end{equation}
Integrating the equation over any eight-sphere gives
\begin{equation}
m\int_{s^8} \ast(dA^{(1)}+mB^{(2)}) =0 \,.
\end{equation}
If however there is a D-particle carrying R-R charge inside the
eight-sphere, this would imply that its flux vanished. To avoid this
conclusion, it was argued that fundamental strings must begin or end
on the D-particle. This adds a source term
$\pm n (2\pi \alpha^\prime)^{-1} \delta^{(8)}(x)$ to eq.~(\ref{msg}),
which integrates to give a D0-brane charge\footnote{ We follow the
convention in \cite{polstrom}, where the coupling of the string to
$B^{(2)}$ is $(2\pi\alpha')^{-1} \int B^{(2)}$.}
\be
\mu_0 = \pm {n \over 2\pi\alpha^\prime m} \,.
\ee
In \cite{polstrom} $m$ was taken to be equal to $\pm\mu_8$, the unit
of D8-brane charge. This reproduces the known relation
\be
\mu_0 \mu_8 = {1\over 2\pi\alpha^\prime} \,
\ee
for $n=1$. In general, the number of fundamental strings is therefore
$n=|m|/\mu_8$; we shall choose the convention that the string is
oriented towards the D-particle if $m>0$, and away from it if $m<0$.

However the value of $m$ on either side of the D8-brane is actually
$\pm \mu_8/2$, as the jump in $m$ across a single
D8-brane is $|\Delta m | = \mu_{8}$ \cite{polwit}. For a single
D8-brane, the
D0-brane is therefore in a region of space with $m=\pm \mu_8/2$, and
the appropriate source term would be
$(4\pi \alpha^\prime)^{-1} \delta^{(8)}(x)$, corresponding to
a string with half a fundamental charge (or tension).

The overall orientation of the strings is fixed uniquely by
supersymmetry. The D8-brane breaks half the supersymmetry of type IIA
string theory, and the D0-brane further breaks half of that, leaving
one quarter of the original supersymmetry (eight supersymmetries). A
string can be added without breaking any further supersymmetry only if
it is perpendicular to the D8-brane and oriented in a particular way
(for the $\overline{\mbox{D}8}$-brane the orientation is opposite).
This fixes the configurations uniquely.

For a D0-brane on the left hand side of a D8-brane where $m=-\mu_8/2$,
the half string stretches between the D8-brane and the D0-brane and it
is oriented in such a way as to go into the D0-brane. On the right hand
side we have $m=\mu_8/2$, and the half string comes out of the
D0-brane and stretches between the D0- and the D8-brane. (This follows
from the fact that the string is oriented the same way on both sides.)
As the D0-brane crosses the D8-brane, it takes its half-string with
it (which goes into the D0-brane, say). At the same time, a fundamental
(whole) string is created which combines with the half-string to give
a half-string of the opposite orientation ({\it i.e.} a half-string
which comes out of the D0-brane); we therefore see that the string
creation is necessary in order to preserve the supersymmetric
orientation of the half-string.

\begin{figure}[htb]
\epsfysize=3cm
\centerline{\epsffile{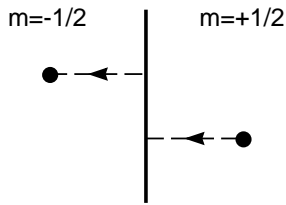}}
\caption{A D0-brane crossing a D8-brane. The dashed line represents
a half-string. The value of $m$ is given in units of $\mu_8$.}
\end{figure}

As half-strings do not seem to be physical, one should really consider
the situation with two D8-branes, where all the strings involved are
real. The analogous process is shown in figure~3, where it is assumed
that $m=-\mu_8$ on the left, $m=0$ in the middle, and $m=+\mu_8$ on
the right. Thus the force vanishes both on the right and on the left.
In the middle the repulsive NS-NS forces due to the two D8-branes
cancel, giving again a vanishing force.

\begin{figure}[htb]
\epsfysize=3.5cm
\centerline{\epsffile{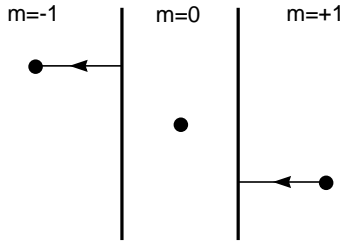}}
\caption{A D0-brane crossing two D8-branes.}
\end{figure}
\smallskip

This analysis does not extend directly to the other (non-compact)
ND$=8$ configurations. For example the perpendicular D1-D7 system in
type IIB string theory exhibits the same amplitude as our D0-D8
system, but the analogous consistency condition due to the equation of
motion for $B^{(2)}$ in massless type IIB supergravity does not
require the introduction of fundamental strings. In the compact case,
however, the relevant low-energy effective theory is nine-dimensional
massive supergravity \cite{brgpt}, and the same result is
obtained. This is also clear by T-duality.

\section{Type IA D-Particles}
\setcounter{equation}{0}

Type IA string theory is a nine-dimensional string theory which is
T-dual to type I theory. It consists of two $\Omega 8$-planes at
$x^9=0$ and $x^9=\pi R_{IA}$, and $16$ D8-branes and their images.
Consider first a D-particle in the presence of a
single $\Omega 8$-plane located at $x^9=0$. (For the time being,
we shall ignore the second orientifold plane and the D8-branes.)
If the D-particle is located away from the $8$-plane, say at $x^9 = R$,
invariance under the combination of world-sheet parity and
reflection along
$x^9$ requires the presence of an image $D$-particle at $x^9=-R$.
The $R$-dependent contribution to the one-loop vacuum energy of this
configuration is given by the one-loop amplitude for the string
stretched between the D-particle and its image
\be
 \A = 2V_1 \int {dk_0 \over 2\pi}  \int_0^\infty {dt\over 2t} \Tr \bigg[
        e^{-2\pi\alpha^\prime t(k_0^2 + M^2)}
        (-1)^{F_S}  {1\over 2}(1+(-1)^F){1\over 2} (1+\Omega I_9)
       \bigg] \; ,
\label{openampl}
\ee
where $\Omega$ denotes world-sheet parity, $I_9$ denotes reflection
along $x^9$,
\be
 M^2 = {(2R)^2\over 4\pi^2\alpha^{\prime 2}} + {1\over\alpha^\prime}
            \sum \mbox{oscillators} + {a\over \alpha'} \,,
\ee
and $a=-1/2$ for the NS-sector, and $a=0$ in the R-sector.
The moding of all the oscillators is as in in (\ref{NNmoding}).
This amplitude can be thought of as the sum of half an annulus
amplitude plus
half a M\"obius strip amplitude. The factor of $2$ outside comes
from exchanging the two ends of the string in the annulus case,
and from a net of $+2$ even minus odd Chan-Paton factors in the
M\"obius strip case.

There are eight potential contributions to the amplitude,
depending on the fermion moding (R or NS), inclusion or not
of $(-1)^F$, and inclusion or not of $\Omega I_9$. The R-sector
has fermion zero-modes in all directions, so the R$(-1)^F$ trace
vanishes.
The action of $\Omega$ on the open string modes is given by
\begin{equation}
 \Omega : \qquad
  \begin{array}{lcl}
   \alpha_n^0 &\longrightarrow & (-1)^n \alpha_n^0 \\
   \alpha_n^{1,\ldots, 9} & \longrightarrow & -(-1)^n
\alpha_n^{1,\ldots, 9}\,,
  \end{array}
\end{equation}
where there is an extra minus sign for the DD directions compared to
the NN direction $x^0$. The action on the fermionic and the ghosts
modes is similar. Together with the obvious action of $I_9$ this
becomes
\be
 \Omega I_9 : \qquad
  \begin{array}{lcl}
   \alpha_n^{0,9} &\longrightarrow & (-1)^n \alpha_n^{0,9} \\
   \alpha_n^{1,\ldots, 8} & \longrightarrow & -(-1)^n
\alpha_n^{1,\ldots, 8}\,.
  \end{array}
\ee
In particular, the action on the fermionic zero-modes can be
represented by
\be
 (\Omega I_9)_0 = \psi_0^1 \cdots \psi_0^8 \,.
\ee
As a consequence, the R$(\Omega I_9)$ trace vanishes, but the
R$((-1)^F\Omega I_9)$ trace gives a non-vanishing contribution,
and the final result  for the potential is
\begin{eqnarray}
V_{D0-\Omega 8}(R)  &=& -\int {dt\over 4t} (8\pi^2\alpha^\prime t)^{-1/2}
         e^{-4R^2 t/(2\pi\alpha^\prime)}  \times \nonumber \\
& & \qquad \qquad \qquad \times
      \left[ {f_3^8(q) - f_4^8(q) - f_2^8(q) \over f_1^8(q)}
     + {- f_4^8(iq) + f_3^8(iq) \pm f_2^8(iq) \over f_2^8(iq)/16}\right]
\nonumber \\
 &=& 8 T_0 R  \Big[ 1 \pm 1 \Big] \,,
\label{08plane}
\end{eqnarray}
where the first three terms correspond to NS, NS$(-1)^F$ and R,
respectively, and the second three terms correspond to
NS$(\Omega I_9)$, NS$((-1)^F\Omega I_9)$ and R$((-1)^F\Omega I_9)$,
respectively. The first three terms cancel due to the
abstruse identity. The sign of the last term corresponds to the choice
of the action of $(-1)^F$ on the zero mode part of the R-sector. A
modular
transformation relates this term to the closed string R-R sector
(Appendix B); its sign corresponds thus to the sign of the R-R
charge, and therefore to whether we are considering an
$\Omega 8$-plane or an $\overline{\Omega 8}$-plane.

The $\Omega 8$-plane is a source of $-16$ units of D8-brane
charge. By symmetry $m=-8\mu_8$ on one side of the $8$-plane, and
$m=+8\mu_8$ on the other side (for the $\overline{\Omega 8}$-plane the
values are reversed). The discussion in section~3 therefore implies
that $8$ fundamental strings must end on each of the D-particles,
with appropriate orientation, as in section~3.
\begin{figure}[htb]
\epsfysize=3.5cm
\centerline{\epsffile{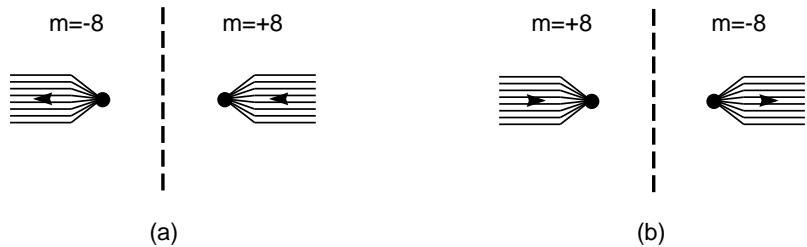}}
\caption{ D0-brane and  image in the presence of  (a) an $\Omega 8$-plane
and (b) an $\overline{\Omega 8}$-plane.}
\end{figure}

\section{ND$=8$-system with relative velocity}
\setcounter{equation}{0}

In this section we shall analyze the situation when there is a
relative velocity between the two branes, where again ND$=8$. For
simplicity we shall refer to this system as the D0-D8-brane system,
but everything we shall say is also true for all ND$=8$
configurations.

There are two different cases to be considered: the D0-brane can move
in a direction parallel to the D8-brane, or in a direction transverse
to the D8-brane. In the first case, the analysis of section~2 is
unchanged, and the potential does not get any velocity dependent
corrections. (This is as usual for the case when two D-branes move in
a relative world-volume  direction.) In the second case, however,
there is a velocity dependent correction. Indeed, we shall find that
for any non-zero transverse velocity the usual calculation of the
phase shift diverges; this divergence is due to the
R$(-1)^F$ sector.

The analysis can be done as in \cite{bac,gil}. In the present case
there is no impact parameter as there is only one transverse
direction, and the phase shift is given by
\be
 \A_{D0-D8} (v)= \int {dt\over 4\pi t}
\frac{\Theta_{1}'(0,it)}{\Theta_{1}(\nu
t,it)}
f_4^{-8} (q)\bigg[ \frac{\Theta_{3}(\nu
t,it)}{\Theta_{3}(0,it)}f_2^8(q) -
\frac{\Theta_{2}(\nu t, it)}{\Theta_{2}(0,it)} f_3^8(q)
+ J_{4} f_4^8(q) \bigg] \,,
\label{08v}
\ee
where $v=\tanh (\nu)$ is the velocity, and $J_4$ will be explained
shortly.
The three terms in the bracket correspond to the contributions from
the NS, R and R$(-1)^F$ sectors, respectively. The  dependence on the
distance $R$ can be recovered from the phase shift by identifying
$R=v\tau$,
where  $\tau$ is the world-line time.

The contribution of  the R$(-1)^F$ sector is unusual. In the static
case, the fermions in this sector have a NN and a DD boundary
condition in the time and the transverse direction and are hence
integer moded. Together with the $(-1)^F$ operator this gives a
fermionic zero-mode in the determinants. (Equivalently, this can be
seen from the fact that the trace over the ground states gives
$(1-1)=0$.)  However, the bosonic superghosts in this sector also have
zero-modes, and the two contributions cancel for zero-velocity to give
the result in eq.~(\ref{08amp}). Once the D0-brane has a velocity in
the transverse direction, the fermions in the time and transverse
directions no longer have zero-modes, but  the superghosts still do;
this therefore gives a divergent result
\be
J_{4}= \infty \times (\frac{\Theta_{1}'(0,it)}
{\Theta_{1}(\nu t,it)})^{-1}\,.
\ee
In the adiabatic approximation, only the leading velocity correction
from the bosonic contribution is considered, and the static potential
(with a changing distance) is recovered. Beyond this approximation the
potential diverges.

If we ignore the contribution from the R$(-1)^F$ sector, we can
compute the velocity dependent potential from the phase shift. We find
at long distances
\be
V_{long}(R,v)=-\frac{1}{2}T_{0}R(1-\frac{1}{2}\frac{v^2}{1-v^2})\,,
\ee
and in the short distance limit (to leading order in the velocity)
\be
V_{short} (R,v) =-\frac{1}{2}T_{0} R -\frac{1}{16}T_{0}\frac{v^2}{R^3}\,.
\ee
This agrees with the result in \cite{df}.

If the D0-brane is between two D8-branes, the two divergences from
the two D8-branes cancel. This will always happen when
the D0-brane is in a region of spacetime where $m=0$. For example,
the divergence is absent when the cosmological constant induced
by the D8-branes is canceled by an $\Omega 8$-plane; in particular,
this is the case in the $SO(16)\times SO(16)$ type IA theory.

On the other hand, if $m\neq 0$, the above analysis seems to suggest
that the D0-branes cannot move in the transverse direction. For
example, let us consider type IIA string theory in the background
of a D8-brane (at infinity). This defines a ten-dimensional theory
whose low energy effective action is massive type IIA supergravity. If
we consider the strong coupling limit of this theory, we might think
that an eleventh dimension will open up and that we will get an
eleven-dimensional theory with a cosmological constant. The
low-energy theory of this theory would be eleven-dimensional
supergravity with a cosmological constant, which does not exist
\cite{Deser}.
On the other hand, the idea of matrix theory \cite{bfss} is
that in the strong coupling limit of type IIA string theory, everything
is made out of  D0-branes. Our analysis therefore suggests that
none of the states can move in the transverse $x^9$ direction, and
that  the resulting theory  is again  ten-dimensional.

\section{World-Line Theories and Matrix Models}
\setcounter{equation}{0}

We can also study the D0-D8 system from the point of view of the
world-line theory of the D-particle. For $R\ll\sqrt{\alpha^\prime}$
this theory is approximately described by the massless open string
modes, where we only keep the lowest derivatives. The resulting theory
is supersymmetric quantum mechanics with $8$ supercharges and
$Spin(8)$ global R-symmetry. The theory was discussed on general
grounds in \cite{bssil}. The analysis of the short distance degrees of
freedom reveals that there is one complex fermion from the
$(0-8)$-string, and the relevant quantum mechanics was written down in
\cite{df,ks,bssil}. However, this description misses a tree-level linear
potential which is due to the R$(-1)^F$-term.  This
potential guarantees that the theory is anomaly free; each complex
fermion leads at one-loop to a  linear potential and a
Chern-Simons term whose coefficients are $\half$, but the
tree-level linear
potential (and the tree-level Chern-Simons term required by
supersymmetry)
has coefficient $\pm\half$ (for each D8-brane), and this
guarantees that the overall coefficient  is always integral.

In the notation of \cite{bssil} the relevant part of the Lagrangian
is then  (taking $2\pi \alpha' = 1$)
\be
{\cal L}=\int dt f(\phi) \dot{\phi}^{2}-if(\phi)\lambda_{a}
\dot{\lambda}_{a} \pm
\frac{1}{2}(\phi+A_{0})-i\bar{\chi}\dot{\chi}-
\bar{\chi}(\phi+A_{0})\chi\,,
\label{l08}
\ee
where $\phi$ denotes the $x^9$ position of  the D0-brane,
$\lambda_a$ is its superpartner (an $8$-component spinor),
$A_0$ is the world-volume gauge field, and $\chi$ is the complex
fermion of the D0-D8 string.
The supersymmetries and gauge transformation of this Lagrangian
were given in \cite{bssil}. The fact that it is impossible to move the
D0-brane along $x^9$ should be  reflected here by the fact that one
cannot give $\phi$ a time dependent expectation value.

We can similarly consider $N$ D0-branes in the presence of a
D8-brane.  The corresponding world-line theory is described  by a
supersymmetric $N\times N$ matrix quantum mechanics with 8
supersymmetries.  Following the idea of \cite{bfss}, the large $N$
limit of  this matrix model should describe the strong coupling
behavior  of type IIA string theory with a non-trivial D8-brane
background. As argued previously,  this should represent
a ten-dimensional, rather than an eleven-dimensional, theory.
\smallskip

The case of a D0-brane and an $\Omega 8$-plane is similar. The
effect of the $\Omega 8$-plane is to induce an additional
linear potential $\pm 8(\phi +A_{0})$. The heterotic matrix theory
corresponds to the situation where there are additional
D8-branes. When eight D8-branes and their images are localized at the
$\Omega$8-plane,  the linear potentials due to the D8-branes and
the orientifold-plane cancel and the world volume theory reduces to
that found in \cite{df,ks}.

In this case, as discussed in the previous section, the D0-brane can
move in the transverse direction (with respect to the D8-branes and the
$\Omega 8$-plane), as $m=0$ everywhere. However, when some of the
D8-branes are not localized at the orientifold plane, there are
regions in space where $m\neq 0$. In such a region there are fundamental
strings between  the D0-brane and some of the D8-branes, and
the D0-branes cannot move non-adiabatically in the transverse direction.
The corresponding tree-level potential depends on $m$, and the
world-line theories appropriate for regions of different $m$ are
therefore different.

\section{Conclusions}
\setcounter{equation}{0}

In this paper we have discussed configurations of branes with
ND$=8$. These configurations have peculiar properties. The one-loop
open string calculation reveals that there is a R-R interaction
between them even though they do not carry the same charge. Furthermore
they cannot move non-adiabatically in the direction transverse to both
of them, and when they cross each other a fundamental string is
created that is stretched between them. The ND$=8$ system  is U-dual
to the system described in \cite{hw}, and the creation of a fundamental
string is U-dual to the creation of a D3-brane when a D5-brane crosses a
NS-$5$-brane.

We have suggested that the inability to move in the transverse
direction is related to the fact that there does not exist a
eleven-dimensional supergravity theory with a cosmological constant
(which would be induced by nine-branes).
\smallskip

We have shown that a system of two branes with ND$=8$ either
has a vanishing force, or a force which is twice the NS-NS force.
The same also holds for the system of a D$p$-brane and an
$\Omega (8-p)$-plane. If we start with a system where the force
vanishes, then the string creation process guarantees that the force
remains zero after the branes have crossed. For the special case of an
isolated D0-D8 system, further constraints imply that the force always
vanishes. Our results are therefore consistent with the previous
analysis of these systems \cite{df,ks,kimray} for the case where eight
D8-branes and their images are at each $\Omega 8$-plane; the results
differ, however, for other configurations.
\medskip

There are many things that need to be understood further.
\begin{list}{(\roman{enumi})}{\usecounter{enumi}}
\item The precise finite velocity dynamics and its implications need
to be understood better.

\item For systems with ND$=8$ other than the D0-D8-system, the nature
of the interaction which is due to the R-R sector is rather unclear.
In particular, the corresponding supergravity theories do not seem to
require that fundamental strings have to be present for consistency.
On the other hand, the corresponding branes {\it can} support
world-volume charges.

Let us take the D1-D7 system as an example. On each of the branes
there is a coupling of the R-R charge of the other brane to the
world-volume gauge potential, and also to a pull back of the
$B^{NS}_{\mu \nu}$ field to the world-volume of the brane. There is
also a coupling in the supergravity of the form $\chi dB^{NS} dB^{R}$,
where $\chi$ is the axion and $B$ is a two-form potential that couples
to the D-string or the elementary string. A combination of these may
be responsible for the extra interaction.

\item The results of this paper may have implications for the
description of the duality between type IA and the heterotic string
theory. For instance the creation of a string as a D8-brane crosses a
D0-brane which is stuck at the $\Omega$8-plane may be related to
the momentum modes which appear on the heterotic side at a point of
enhanced gauge symmetry \cite{polwit,psj}.

\item More speculative is the connection to the heterotic matrix
theory. It is known that upon compactification on $S^1$, turning on
Wilson lines seems to create an anomaly. It was suggested in
\cite{horava} that this may be canceled with an anomaly inflow from
the bulk. It was shown in \cite{bdg} that the anomaly inflow has to do
with the creation of fundamental strings, which, as we have seen, is
connected to an extra term in the interaction of D0-branes and
D8-branes (or an  $\Omega 8$-plane). Taking this extra interaction
into account may help to resolve the problem of compactifications.
\end{list}

\noindent These issues are currently under study.

\section*{Acknowledgments}

We thank  S. Deser, A. Hanany, J. Polchinski, S.J. Rey, C. Vafa,
E. Witten and B. Zwiebach for useful discussions.

O.B. is supported in part by the NSF under grants PHY-93-15811
and PHY-92-18167, and  M.R.G. is supported by a
NATO-Fellowship and in part by NSF grant PHY-92-18167.

\appendix
\section*{Appendix: Closed String Calculations}

\section{D0-D8}
\renewcommand{\theequation}{A.\arabic{equation}}
\setcounter{equation}{0}

In this section we shall explain how the results can also be
obtained by performing a closed string calculation. We shall work
covariantly, and we shall use the same conventions as in
\cite{Billo} (see also \cite{CalKle}).

We shall first describe boundary states which satisfy Neumann boundary
conditions for $x^\mu$ with $\mu=0, \ldots, p$, and Dirichlet boundary
conditions for $x^i$ with $i=p+1, \ldots, 9$. We shall then also
consider the boundary states which can be obtained from these by
Lorentz transformations. We shall work in $d=10$ spacetime
dimensions, and we denote by $d^\perp$ the number of transverse
dimensions, {\it i.e.} $d^\perp=9-p$.

Let us first consider boundary states which are localized at
the transverse position $y^i$. These can be
described as linear combinations of coherent states of the form
\be
|Bp,y^i,\eta\rangle = |Bp,y^i\rangle_b \, |Bp,\eta\rangle_f \,
|B, \eta \rangle_g \,,
\ee
where the subscripts $b$, $f$ and $g$ denote the bosonic, fermionic
and ghost sector, respectively. The component in the bosonic sector
$|Bp,y^i\rangle_b$ is
\be
\label{Bpbos}
(2 \pi \sqrt{\alpha'})^{d^\perp}
\prod_{i=p+1}^{9} \delta(q^i - y^i) \,
\exp\left\{ \sum_{n=1}^{\infty} {1 \over n} \left[
  - \eta_{\mu\nu}\alpha_{-n}^{\mu} \tilde\alpha_{-n}^{\nu}
+ \alpha_{-n}^{i} \tilde\alpha_{-n}^{i}\right]
\right\} |0\rangle\,,
\ee
where $\eta_{\mu\nu} = \mbox{diag}(-1,1,\ldots,1)$, and $|0\rangle$ is
the zero-momentum ground state. Here the sum over $\mu$ and $\nu$
runs over $0,\ldots, p$, and the sum over $i$ is from $p+1$ to $9$.
The other two components
depend on whether we are considering the NS-NS sector or the R-R
sector. In the first case, the fermionic state $|Bp,\eta\rangle_f$ is
\be
|Bp,\eta\rangle_{NSNS} =
\exp\left\{i \eta \sum_{r>0} \left[
 -\eta_{\mu\nu}\psi_{-r}^{\mu} \tilde\psi_{-r}^{\nu}
+  \psi_{-r}^{i} \tilde\psi_{-r}^{i}\right]
\right\} |0\rangle_{NSNS}\,,
\ee
where $|0\rangle_{NSNS}$ is the NS-NS ground state, and in the
second case
\be
|Bp,\eta\rangle_{RR} =
\exp\left\{i \eta \sum_{m=1}^{\infty} \left[
 -\eta_{\mu\nu}\psi_{-m}^{\nu} \tilde{\psi}_{-m}^{\nu}
+  \psi_{-m}^{i} \tilde{\psi}_{-m}^{i}\right]
\right\} |p,\eta\rangle_{RR}^0\,,
\ee
where $|p,\eta\rangle_{RR}^0$ is the R-R ground state
satisfying
\begin{equation}
  \psi^\mu_\mp |p,\pm\rangle_{RR}^0 = \psi^i_\pm |p,\pm\rangle_{RR}^0 = 0
  \,, \qquad
\psi^\alpha_{\pm} \equiv {1 \over \sqrt{2}} \left( \psi_0^{\alpha}
\pm i \tilde{\psi}_0^{\alpha} \right) \,.
\end{equation}
The coherent state in the ghost sector $|B \eta \rangle_g$ does not
depend on $p$, and is as given in \cite{PolCai}. As explained in
\cite{bergmangaberd} (see also \cite{PolCai,Billo}), invariance under
the GSO projection (and the consistency with the open string sector)
requires that the physical $D$-brane states are linear combinations of
these states
\be
|Dp, y^i \rangle = |Dp,y^i\rangle_{NSNS} + |Dp,y^i\rangle_{RR} \,,
\ee
where
\be
|Dp,y^i\rangle_{NSNS} ={\Nu_{NSNS}\over 2}\left(|Bp,y^i,+\rangle_{NSNS} -
|Bp,y^i,-\rangle_{NSNS} \right)
\ee
is the component in the NS-NS sector, and
\be
|Dp,y^i\rangle_{RR} ={\Nu_{RR}\over 2}\left( |Bp,y^i,+\rangle_{RR} +
|Bp,y^i,-\rangle_{RR} \right)
\ee
is the component in the R-R sector. The GSO condition also implies
that $p$ is even for type IIA, odd for type IIB, and $p=1,5,9$ in type
I. We normalize the ground states so that $\mbox{}_{NSNS}\langle
0|0\rangle_{NSNS}=1$,
and
\be
\label{Rnorm}
 \mbox{}_{RR}^{\quad 0} \langle p^\prime,\eta^\prime \ket{p,\eta}_{RR}^0
  = \delta_{p,p'} \delta_{\eta,\eta'}
  + \delta_{|p-p'|,10} \delta_{\eta,-\eta'} \,.
\ee
The normalization constants $\Nu_{NSNS}$ and $\Nu_{RR}$ are then
determined up to a sign by consistency with the open string
calculation. As the overall sign is irrelevant, there
exist two different solutions which differ by the relative sign, and
they correspond to the brane and the anti-brane solution,
respectively.
\smallskip

The amplitude to propagate from one D-brane state to another is
given, up to a normalization, by
\begin{equation}
 \A_{Dp-Dp^\prime} = \int_0^\infty {d\tau\over \tau} \bra{Dp,y}
    e^{-\pi\tau(L_0 +  \tilde{L}_0)} \ket{Dp^\prime,y^\prime} \; .
\end{equation}
To fix the normalization constants we shall first determine the
amplitude between two stationary $p$-branes. Using the conventions
of \cite{Billo} for the normalization of the amplitude (including
the same normalization factor of $(2\alpha')^{-d/2}/\pi$)
we find for the contribution in the NS-NS sector
\be
\A_{Dp-Dp}^{NSNS} =  {\Nu_{NSNS}^2 \over 2}
{V_{p+1} \over (8\pi^2 \alpha')^{(p+1)/2}}
\int_{0}^{\infty} d \tau \tau^{-4+(p-1)/2}
e^{-R^2 / 2\pi\alpha'\tau}
{f_3^8(r) - f_4^8(r) \over f_1^8(r)} \,,
\ee
where $R$ is the transverse distance between the two $p$-branes, and
$r=e^{-\pi\tau}$; in the R-R sector we get
\be
\A_{Dp-Dp}^{RR} = {\Nu_{RR}^2 \over 32}
{V_{p+1} \over (8\pi^2 \alpha')^{(p+1)/2}}
\int_{0}^{\infty} d \tau \tau^{-4+(p-1)/2}
e^{-R^2 / 2\pi\alpha'\tau}
{f_2^8(r) \over f_1^8(r)} \,.
\ee
Here the functions $f_i$ are defined as
\be
\label{fi}
\begin{array}{lcl}
f_1(q) & = & {\displaystyle
q^{1/12} \prod_{n=1}^{\infty} ( 1- q^{2n}) } \\
f_2(q) & = & {\displaystyle
\sqrt{2} q^{1/12} \prod_{n=1}^{\infty} (1 + q^{2n}) } \\
f_3(q) & = & {\displaystyle
q^{-1/24} \prod_{n=1}^{\infty} (1 + q^{2n-1}) } \\
f_4(q) & = & {\displaystyle
q^{-1/24} \prod_{n=1}^{\infty} (1 - q^{2n-1})\,.}
\end{array}
\ee
It follows that the total amplitude
$\A_{Dp-Dp}^{NSNS} + \A_{Dp-Dp}^{RR}$
vanishes provided that $\Nu_{RR} = 4 i \Nu_{NSNS}$. To fix the remaining
overall constant, we substitute $t=1 / \tau$, and perform a modular
transformation to obtain
\be
\A_{Dp-Dp}^{NSNS} = {\Nu_{NSNS}^2 \over 2}
{V_{p+1} \over (8\pi^2 \alpha')^{(p+1)/2}}
\int_{0}^{\infty} {dt \over t} t^{-(p+1)/2}
e^{-R^2 t / 2\pi\alpha'}
{f_3^8(q) - f_2^8(q) \over f_1^8(q)} \,,
\ee
where $q=e^{-\pi t}$. This agrees with the open string calculation
provided we set $\Nu_{NSNS}=1$.

Now that all normalization constants are determined, we can calculate
the amplitude for a stationary D0-D8 configuration. In
the NS-NS sector we find by a similar calculation
\be
\label{08NSNS}
\A_{D0-D8}^{NSNS} =  {V_{1} \over (8\pi^2 \alpha')^{1/2}}
{1 \over 2}
\int_{0}^{\infty} d \tau \tau^{-1/2}
e^{-R^2 / 2\pi\alpha'\tau}
{f_4^8(r) - f_3^8(r) \over f_2^8(r)} \,,
\ee
where again $r=e^{-\pi\tau}$. Using the abstruse identity it follows
that the oscillator contribution (the ratio of the $f$-functions)
equals $-1$. Substituting $t=1 / \tau$ and performing a modular
transformation we then find
\begin{eqnarray}
\A_{D0-D8}^{NSNS} & = & {1 \over 2}
{V_{1} \over (8\pi^2 \alpha')^{1/2}}
\int_{0}^{\infty} {dt \over t} t^{-1/2} e^{-R^2 t / 2\pi\alpha'}
(-1) \nonumber \\
& = & - {1 \over 2} \, {V_1 \over (2 \pi \alpha')} \, R \,
{\Gamma(-1/2) \over 2 \sqrt{\pi} }
=  {1 \over 2} \, {V_1 \over (2 \pi \alpha')} \, R \,,
\end{eqnarray}
which agrees with the open string calculation, and
reproduces the crucial normalization factor. Superficially, the
contribution from the R-R sector vanishes since the R-R ground states
only contribute for $p=p'$ and $|p-p'|=10$ as follows from
(\ref{Rnorm}). However, the calculation is more subtle as the
contribution from the superghost ground state has to be taken into
account. Indeed, the contribution of the superghosts to the amplitude
$\bra{B0,y_0,\eta} e^{-\pi\tau(L_0 + \tilde{L}_0)} \ket{B8,y_8,-\eta}$ is
\be
\prod_{n} (1 - r^{2n})^{-2} \, ,
\ee
and thus their ground state gives a divergent term ($n=0$), which
can potentially cancel the vanishing R-R ground state amplitude. In
fact,  it is clear from the gauge fixing condition for the light cone
gauge, that there are only eight physical zero modes, and the
light-cone  calculation (see for example \cite{bergmangaberd}) then
implies that there is a non-vanishing contribution from the R-R sector
for $|p-p'|=8$. This contribution cancels precisely the NS-NS
contribution if one boundary state is a brane, and one an anti-brane,
and doubles the NS-NS contribution in the case where both boundary
states are branes or anti-branes.
\bigskip

Next we shall consider the boundary states that can be obtained from
the stationary boundary states by the application of a Lorentz
transformation. We want to analyze first the case, where the Lorentz
transformation corresponds to a pure space-rotation; in this case, we
want to show that a rotation by $\pi$ in a plane which is spanned by
one Neumann and one Dirichlet direction turns a $p$-brane into an
anti-$p$-brane, whereas a rotation by $\pi$ in a plane spanned by two
Neumann or two Dirichlet directions leaves the $p$-brane invariant.

Let us recall \cite{GSW} that the generator of the Lorentz
transformation is given as
\be
J^{\mu\nu} = l^{\mu\nu} + E^{\mu\nu} + K^{\mu\nu} \,,
\ee
where $l^{\mu\nu} = q^\mu p^\nu - q^\nu p^\mu$,
\be
E^{\mu\nu} = - i \sum_{n=1}^{\infty} {1 \over n}
\left( \alpha^\mu_{-n} \alpha^\nu_{n}
- \alpha^\nu_{-n} \alpha^\mu_{n}
+ \tilde\alpha^\mu_{-n} \tilde\alpha^\nu_{n}
- \tilde\alpha^\nu_{-n} \tilde\alpha^\mu_{n}\right) \,,
\ee
and
\be
K^{\mu\nu} = - i \sum_{r>0}
\left( \psi^\mu_{-r} \psi^\nu_{r}
- \psi^\nu_{-r} \psi^\mu_{r}
+ \tilde\psi^\mu_{-r} \tilde\psi^\nu_{r}
- \tilde\psi^\nu_{-r} \tilde\psi^\mu_{r}\right)
\ee
in the NS-NS sector, and
\be
K^{\mu\nu} =
- {i \over 2}\left( [\psi_0^\mu,\psi_0^\nu]
+ [\tilde{\psi}_0^\mu,\tilde{\psi}_0^\nu] \right)
-i \sum_{m=1}^{\infty}
\left( \psi^\mu_{-m} \psi^\nu_{m}
- \psi^\nu_{-m} \psi^\mu_{m}
+ \tilde{\psi}^\mu_{-m} \tilde{\psi}^\nu_{m}
- \tilde{\psi}^\nu_{-m} \tilde{\psi}^\mu_{m}\right)
\ee
in the R-R sector. A finite Lorentz transformation is then of the form
\be
U(\alpha) = \exp\left(i \theta_{\mu\nu} J^{\mu\nu} \right)
\,.
\ee
We shall consider a spatial rotation in the $i-j$ plane, and this
corresponds to the situation where $\theta_{\mu\nu}=0$ unless
$(\mu,\nu)=(i,j)$. It is then easy to check that
\be
U(\theta_{ij}) \alpha^k_{-n} U(\theta_{ij})^{-1} =
\left\{
\begin{array}{ll}
\alpha^k_{-n} & \hbox{if $k\neq i,j$,} \\
\hbox{cos} (\theta_{ij}) \alpha^i_{-n}
- \hbox{sin}(\theta_{ij}) \alpha^j_{-n}
& \hbox{if $k=i$,} \\
\hbox{cos} (\theta_{ij}) \alpha^j_{-n}
+ \hbox{sin}(\theta_{ij}) \alpha^i_{-n}
& \hbox{if $k=j$.}
\end{array}
\right.
\ee
If we take $\theta_{ij}=\pi$, it then follows that,
apart from the bosonic zero modes which transform in the obvious way,
the bosonic sector and the NS-NS sector of the boundary state are
invariant under $U(\pi)$. Similarly the oscillator part of the R-R sector
is also invariant. The action on the R-R ground state however
depends on the boundary conditions in the plane of rotation $(i,j)$.
The relevant part of $U(\pi)$
is $\exp\left(\pi ( \psi^i_{+} \psi^j_{-} + \psi^i_{-} \psi^j_{+} )
\right)$,
whose action gives
\be
\label{Rground}
\exp\left(\pi ( \psi^i_{+} \psi^j_{-} + \psi^i_{-} \psi^j_{+} ) \right)
  \ket{p,\eta}_{RR}^0  = \left\{
   \begin{array}{rl}
     \ket{p,\eta}_{RR}^0 & \mbox{for (N,N) or (D,D)} \\
      - \ket{p,\eta}_{RR}^0 & \mbox{for (N,D) or (D,N).}
   \end{array}
 \right.
\ee
Such a rotation therefore changes the sign of the R-R component, and
thus transforms a $p$-brane into an anti-$p$-brane and vice
versa.

Next we shall analyze the effect of a boost on the boundary
state. Following the analysis of \cite{Billo}, we find that
a boost in the $k$-th direction with rapidity $\nu$ (where
$k\geq p+1$) transforms the boundary state $|p,y^i,\eta\rangle$ to
\be
|Bp,y^i,v,\eta\rangle = |Bp,y^i,v\rangle_b \, |Bp,\eta,v\rangle_f \,
|B, \eta \rangle_g \,,
\ee
where $|Bp,y^i,v\rangle_b$ is now
\begin{eqnarray}
|Bp,y^i,v\rangle_b & = &
(\alpha')^{d^\perp/2}\, \sqrt{1-v^2}\, \delta(q^k - q^0 v - y^k)
\prod_{i\neq k} \delta(q^i - y^i) \nonumber \\
& & \exp\left\{ \sum_{n=1}^{\infty} {1 \over n} \left[
- \sum_{\nu=1}^{p} \alpha_{-n}^{\nu} \tilde\alpha_{-n}^{\nu}
+ \sum_{i\neq k} \alpha_{-n}^{i} \tilde\alpha_{-n}^{i}\right]\right\}
\nonumber \\
& & \exp\left\{ \sum_{n=1}^{\infty} {1 \over n}
+ \hbox{cosh}(2\nu) \left(\alpha_{-n}^{0} \tilde\alpha_{-n}^{0}
+ \alpha_{-n}^{k} \tilde\alpha_{-n}^{k}\right)
\right\} \nonumber \\
& & \exp\left\{ \sum_{n=1}^{\infty} {1 \over n} \left[
\hbox{sinh}(2\nu) \left(\alpha_{-n}^{0} \tilde\alpha_{-n}^{k}
+ \alpha_{-n}^{k} \tilde\alpha_{-n}^{0}\right)\right]
\right\} |0\rangle\,,
\end{eqnarray}
and $v=\hbox{tanh}(\nu)$ is the velocity.
A similar formula holds for the NS-NS fermions, and for the
non-zero oscillators in the R-R sector. The R-R ground state
on the other hand becomes
\be
\label{RRv}
\ket{p,\pm,v}_{RR}^0 = \hbox{cosh}(\nu) \ket{p,\pm}_{RR}^0
+ \hbox{sinh}(\nu) \psi^0_{\pm} \psi^k_{\mp}\ket{p,\pm}_{RR}^0\,.
\ee

Using these expressions we can evaluate the amplitude for the
configuration of a static $8$-brane, and a $0$-brane with velocity
$\tanh(\nu)$ in the $x^9$ direction. Using the same normalization as
before we obtain in the NS-NS sector
\begin{eqnarray}
\label{08vNSNS}
\A^{NSNS}_{D0-D8}(v) & = & {1 \over 4} {1 \over \hbox{sinh}(\nu)}
\int_{0}^{\infty} d\tau {1 \over f_2^8(r)}
\prod_{n=1}^{\infty} {(1 - r^{2n})^2 \over (1 - e^{2\nu} r^{2n})
(1 - e^{-2\nu}r^{2n})} \nonumber \\
& &
\left(f_4^8(r) \prod_{n=1}^{\infty}
{(1 + e^{2\nu} r^{2n-1}) (1 + e^{-2\nu}r^{2n-1}) \over
(1 + r^{2n})^2} \right. \nonumber \\
& & \left. \qquad
- f_3^8(r) \prod_{n=1}^{\infty}
{(1 - e^{2\nu} r^{2n-1}) (1 - e^{-2\nu}r^{2n-1}) \over
(1 - r^{2n})^2}\right) \,,
\end{eqnarray}
where $r=e^{-\pi\tau}$. In the R-R sector, it follows from
(\ref{RRv}), together with (\ref{Rnorm}) and
\be
\mbox{}_{RR}^{\quad 0}\bra{0,\eta_0} \psi^0_{\pm}  \psi^9_{\mp}
         \ket{8,\eta_8}_{RR}^0 =  - \delta_{\eta_0,-\eta_8} \,,
\ee
that the R-R component of the amplitude vanishes unless
$\eta_8 = -\eta_0$, and the contribution is then proportional to
$\hbox{sinh}(\nu)$. In addition we find that all oscillator
contributions cancel, and the amplitude becomes
\be
\label{08vRR}
\A^{RR}_{D0-D8}(v) = - {SG_0 \over 4}
\int_{0}^{\infty} d\tau \,,
\ee
where we have represented by $SG_0$ the divergent contribution of the
superghost ground state of
$\langle \eta | e^{-\pi\tau(L_0+\tilde{L}_0)} |-\eta\rangle_g$. The
R-R amplitude is independent of the velocity, and thus persists in the
limit $\nu\rightarrow 0$. This is necessary in order to cancel the
$\mbox{sinh}(\nu)^{-1}$ divergence of the NS-NS amplitude in this
limit.

\section{D0-${\bf \Omega 8}$ }
\renewcommand{\theequation}{B.\arabic{equation}}
\setcounter{equation}{0}

Just as the original  boundary states of Polchinski
and Cai \cite{PolCai} can be generalized to boundary states describing
D$p$-branes, one can generalize the crosscap
states. Indeed, it is natural to introduce the concept of a
$p$-crosscap as the state which satisfies the conditions
\begin{eqnarray}
\label{crosscapcon}
(\alpha_n^\mu + e^{i\pi n}\widetilde{\alpha}_{-n}^\mu) \,
\ket{Cp,\eta} =
(\psi_m^\mu + \eta e^{i\pi m}\widetilde{\psi}_{-m}^\mu) \,
\ket{Cp,\eta} & = & 0 \qquad \mu = 0, \ldots ,p, \nonumber \\
(\alpha_n^i - e^{i\pi n}\widetilde{\alpha}_{-n}^i) \,
\ket{Cp,\eta} =
(\psi_m^i - \eta e^{i\pi m}\widetilde{\psi}_{-m}^i)\,
\ket{Cp,\eta} & = & 0 \qquad i = p+1,\ldots, 9\, .
\end{eqnarray}
Up to normalization, the unique solution is
\begin{equation}
|Cp,y^i,\eta\rangle = |Cp,y^i\rangle_b \, |Cp,\eta\rangle_f \,
|C, \eta \rangle_g \,,
\ee
where $|Cp,y^i\rangle_b$ is
\be
(2 \pi \sqrt{\alpha'})^{d^\perp}
\prod_{i=p+1}^{9} \delta(q^i - y^i)
\exp\left\{ \sum_{n=1}^{\infty} {(-1)^n \over n} \left[
  - \eta_{\mu\nu} \alpha_{-n}^{\mu} \tilde\alpha_{-n}^{\nu}
+ \alpha_{-n}^{i} \tilde\alpha_{-n}^{i}\right]
\right\} |0\rangle\,.
\ee
The fermionic contribution in the NS-NS sector is
\be
|Cp,\eta\rangle_{NSNS} =
\exp\left\{i \eta \sum_{r>0} e^{i\pi  r} \left[
 - \eta_{\mu\nu} \psi_{-r}^{\mu} \tilde\psi_{-r}^{\nu}
+ \psi_{-r}^{i} \tilde\psi_{-r}^{i}\right]
\right\} |0\rangle_{NSNS}\,,
\ee
and the contribution in the R-R sector is
\be
|Cp,\eta\rangle_{RR} =
\exp\left\{i \eta \sum_{m=1}^{\infty} (-1)^m \left[
 - \eta_{\mu\nu} \psi_{-m}^{\mu} \tilde{\psi}_{-m}^{\nu}
+ \psi_{-m}^{i} \tilde{\psi}_{-m}^{i}\right]
\right\} |p,\eta\rangle_{RR}^0\,,
\ee
where $|p,\eta\rangle_{R}$ is the same ground state as defined before,
since it satisfies the same equations as before, {\it i.e.}
(\ref{crosscapcon}) with $m=0$. The ghost sector state $|C \eta
\rangle_g$ is independent of $p$, and is given as in \cite{PolCai}.

As before, invariance under the GSO projection and  consistency
with the open string sector restrict the physical orientifold plane
($\Omega$-plane) states to be
\be
\ket{\Omega p} =
{\Nu^C_{NSNS} \over 2} \Big(\ket{Cp,+}_{NSNS} - \ket{Cp,-}_{NSNS}\Big)
\pm
{\Nu^C_{RR} \over 2} \Big(\ket{Cp,+}_{RR} + \ket{Cp,-}_{RR}\Big) \,,
\ee
where the normalization constants are fixed by consistency with
the open string M\"obius strip to be
$\Nu^C_{NSNS} = 2^{p-4}i$ and $\Nu^C_{RR}=2^{p-2} i$.
The choice of sign reflects the fact that there exist
anti-orientifold planes as well as orientifold planes; this is in
complete analogy to the situation for Dirichlet branes.

The stationary amplitude between a D$p$-brane and an $\Omega
p^\prime$-plane
can be calculated in the same way as the amplitude of two D-branes.
The NS-NS contribution is given by (assuming $p^\prime>p$)
\begin{equation}
 \A_{Dp-\Omega p^\prime}^{NSNS}  =  C \int_{0}^{\infty} d \tau
   \tau^{(p^\prime - 9)/2} e^{-R^2 / 2\pi\alpha'\tau}
     {f_3(ir)^{8+p-p^\prime} f_4(ir)^{p^\prime - p}
    - f_4(ir)^{8+p-p^\prime} f_3(ir)^{p^\prime - p}
    \over f_1(ir)^{8+p-p^\prime}f_2(ir)^{p^\prime - p}} \, ,
\end{equation}
and the R-R contribution is
\begin{equation}
 \A_{Dp-\Omega p^\prime}^{RR}  =  C \int_{0}^{\infty} d \tau
   \tau^{(p^\prime - 9)/2} e^{-R^2 / 2\pi\alpha'\tau} \left[
     {f_2^8(ir) \over f_1^8(ir)} \delta_{p,p^\prime}
    + \delta_{p,p^\prime - 8} \right] \, .
\end{equation}
The overall constant $C$ is the same in both cases, and is given by
\begin{equation}
 C = - 2^{p^\prime -5} (8\pi^2 \alpha^\prime)^{(p^\prime -9)/2} V_1 \; .
\end{equation}
For D$0$-$\Omega 8$ this gives
\begin{eqnarray}
 \A_{D0-\Omega 8}^{NSNS}   &=&   8  {V_{1} \over (8\pi^2 \alpha')^{1/2}}
  \int_{0}^{\infty} d \tau \tau^{-1/2}
   e^{-R^2 / 2\pi\alpha'\tau} \nonumber \\
 &=&   - 8 \, {V_1 \over (2 \pi \alpha')} \, R \,,
\label{omega8nsns}
\end{eqnarray}
and the same for the R-R contribution. As in the D0-D8 case, there
are two
choices for the relative sign between the NS-NS and R-R contributions,
one corresponding to an $\Omega 8$-plane and the other to
an $\overline{\Omega 8}$-plane. In one case the total amplitude
vanishes, and in the other it is twice (\ref{omega8nsns}).

Similarly, the amplitude describing a static $\Omega 8$-plane and
a D0-brane moving transverse to it is found
to be
\begin{eqnarray}
\A^{NSNS}_{D0-\Omega 8}(v) & = & 4 {1 \over \hbox{sinh}(v)}
\int_{0}^{\infty} d\tau {1 \over f_2^8(ir)}
\prod_{n=1}^{\infty} {(1 - (ir)^{2n})^2 \over (1 - e^{2v} (ir)^{2n})
(1 - e^{-2v} (ir)^{2n})} \nonumber \\
& &
\left(f_4^8(ir) \prod_{n=1}^{\infty}
{(1 + e^{2v} (ir)^{2n-1}) (1 + e^{-2v} (ir)^{2n-1}) \over
(1 + (ir)^{2n})^2} \right. \nonumber \\
& & \left. \qquad
- f_3^8(ir) \prod_{n=1}^{\infty}
{(1 - e^{2v} (ir)^{2n-1}) (1 - e^{-2v}(ir)^{2n-1}) \over
(1 - (ir)^{2n})^2}\right) \,,
\end{eqnarray}
and the corresponding R-R amplitude is
\be
\A^{RR}_{D0-\Omega 8}(v) = - 4 (SG_0)
\int_{0}^{\infty} d\tau \,,
\ee
where $SG_0$ is again the divergent contribution from the superghost
ground states.

\end{document}